\documentclass[12pt,a4paper]{article}
\usepackage{amssymb}

\title{From supermembrane to super Yang-Mills theory}
\author{
{\sc Shozo Uehara}\footnote{e-mail:
uehara@eken.phys.nagoya-u.ac.jp}~ and
{\sc Satoshi Yamada}\footnote{e-mail:
yamada@eken.phys.nagoya-u.ac.jp}\vspace{4mm}\\
{\sl Department of Physics, Nagoya University}\\
{\sl Chikusa-ku, Nagoya 464-8602, Japan}}
\date{}

\makeatletter

\@addtoreset{equation}{section}
\renewcommand{\thefigure}{\@arabic\c@figure}
\makeatother

\newcommand{\nn}{\nonumber\\}
\newcommand{\ptau}{\partial_\tau}
\newcommand{\psig}{\partial_\sigma}
\newcommand{\prho}{\partial_\rho}
\newcommand{\mtheta}{\mathit{\Theta}}

\newcommand{\p}{\partial}
\newcommand{\scom}[1]{[\,#1\,]_*}

\begin{document}
\maketitle
\vspace{-80mm}
\begin{flushright}
DPNU-04-10\\
hep-th/0405037\\
May 2004
\end{flushright}
\vspace{57mm}

%%%%%%%%%%%%%%%%%%%%%%%%%%%%%%%%%
\begin{abstract}
%%%%%%%%%%%%%%%%%%%%%%%%%%%%%%%%%
We derive $p$+1-dimensional ($p$=1,2) maximally supersymmetric $U(N)$
Yang-Mills theory from the wrapped supermembrane on
$R^{11-p}\times T^{p}$ in the light-cone gauge by using the matrix
regularization.
The elements of the matrices in the super Yang-Mills theory are given
by the Fourier coefficients in the supermembrane theory.
Although our approach never refers to both
D-branes and superstring dualities, we obtain the relations which
exactly represent T-duality.
\end{abstract}

%%%%%%%%%%%%%%%%%%%%%%%%%%%%%%%%%%
\section{Introduction}
%%%%%%%%%%%%%%%%%%%%%%%%%%%%%%%%%%
Supermembrane in eleven dimensions \cite{BST} is believed to play an
important role to understand the fundamental degrees of freedom
in M-theory.
Actually, the matrix-regularized light-cone supermembrane on $R^{11}$
\cite{Hop,dWHN}, which is 0+1-dimensional maximally supersymmetric
$U(N)$ Yang-Mills theory, is conjectured to describe the
light-cone quantized M-theory in the large-$N$ limit \cite{BFSS}.
Furthermore, Susskind suggested that even at finite $N$, the super
Yang-Mills theory describes the $p^{+}=N/R$ sector of discrete
light-cone quantized (DLCQ) M-theory \cite{Sus}.\footnote{In this
paper we use a convention of the light-cone coordinates $x^\pm\equiv
(x^{0}\pm x^{10})/\sqrt2$.
Then, in DLCQ, $x^-$ is compactified on $S^1$ with radius $R$.}
Hence, the super Yang-Mills theory is called Matrix theory, or
M(atrix) theory, by identifying $N$ of the gauge group $U(N)$ with
that of the light-cone momentum $p^{+}=N/R$.
Susskind's conjecture was explained in Refs.\cite{Sei,Sen}, where
Matrix theory is interpreted as the low energy effective theory of $N$
D$0$-branes and the 11th direction is chosen as the longitudinal
direction.
Thus, Matrix theory is looked on either as the matrix-regularized
theory of the light-cone supermembrane on $R^{11}$ or the low energy
effective theory of D0-branes.

{}From the latter viewpoint, which we call the {\it D-brane
viewpoint}, we can easily carry out the toroidal compactification of
(the transverse directions of) Matrix theory.
In fact, through the T-duality prescription \`a la Taylor \cite{Tay},
Matrix theory compactified on a circle is the low energy effective
theory of $N$ D$1$-branes, i.e., the 1+1-dimensional
maximally supersymmetric $U(N)$ Yang-Mills theory.
The super Yang-Mills theory is called matrix string theory,
which is conjectured to describe the light-cone quantized type-IIA
superstring theory in the large-$N$ limit \cite{Mot,DVV}.
It is also conjectured to describe the $p^{+}=N/R$ sector of
DLCQ type-IIA superstring theory even at finite $N$ \cite{Sus}.
This proposal is explained by using the T- and S-dualities with the
9-11 flip of interchanging the role of the 11th and 9th directions
\cite{Mot,DVV}.
Furthermore, it is straightforward to compactify Matrix theory on
$T^p$ ($p\leq 3$) and the resulting theory is the low energy effective
theory of $N$ D$p$-branes, i.e., the $p$+1-dimensional maximally
supersymmetric $U(N)$ Yang-Mills theory \cite{Sus2,GRT,FHRS}.

On the other hand, from the former point of view, which we call the
{\it membrane viewpoint}, it is not so straightforward to compactify
Matrix theory on a torus.
Actually, it is only recently that matrix string theory has deduced
via matrix regularization of the light-cone wrapped supermembrane on
$R^{10}\times S^1$ \cite{SY,Ced,UY3}.\footnote{See
Refs.\cite{HI,Shi,DW,CL} for recent other approaches to the membrane
theory.}
Furthermore, Matrix theory compactified on higher dimensional torus
from the membrane viewpoint is not known yet.

The purpose of this paper is to drive systematically Matrix
theory compactified on $T^p$ ($p=1,2$) {\it from the membrane
viewpoint}.
Namely, we derive $p$+1-dimensional maximally supersymmetric $U(N)$
Yang-Mills theory through the matrix regularization of the wrapped
supermembrane on  $R^{11-p}\times T^p$.
In the case of $p=1$, it was already presented in
Refs.\cite{SY,Ced,UY3}. However, our explanation in this paper is
improved and is successfully extended to the case of $p=2$.
Furthermore, we naturally obtain the relations which exactly agree
with those inferred from T-duality. We also derive the dimensionless
gauge coupling constant in the super Yang-Mills theory and find that
it agrees with that obtained in  Ref.\cite{FHRS}, where they derived
it from the D-brane viewpoint. Thus, this gives a consistency check of
two kinds of approaches in the toroidal compactification of Matrix
theory.

The plan of this paper is as follows.
We start by giving the mode expansions of eleven-dimensional
supermembrane in the light-cone gauge.
In section \ref{S:0+1}, just as a warm-up, we present the derivation
of 0+1-dimensional maximally supersymmetric $U(N)$ Yang-Mills theory
through the matrix regularization of the supermembrane on $R^{11}$.
In section \ref{S:1+1}, we give a simpler derivation of 1+1-dimensional
maximally supersymmetric $U(N)$ Yang-Mills theory from the wrapped
supermembrane on $R^{10}\times S^1$ \cite{SY,Ced,UY3}. We also derive
a certain relation which implies T-duality.
In section \ref{S:2+1}, we go on to 2+1-dimensional maximally
supersymmetric $U(N)$ Yang-Mills theory with the wrapped supermembrane
on $R^{9}\times T^2$. Final section is devoted to summary and
discussion.

%%%%%%%%%%%%%%%%%%%%%%%%%%%%%%%%%%
\section{Supermembrane in the light-cone gauge}
%%%%%%%%%%%%%%%%%%%%%%%%%%%%%%%%%%
Our starting point is the action of the eleven-dimensional
supermembrane in the light-cone gauge,\footnote{Precisely speaking,
when the membrane has a non-trivial space-sheet topology, we need to
impose the global constraints to the action (\ref{MLCgauge})
\cite{UY2}. For simplicity, such constraints are ignored in
this paper.}
(Here we just write it only with the bosonic degrees of
freedom. Fermions are straightforwardly included.)
\begin{eqnarray}
  S&=&\frac{LT}{2}\int d\tau \int_0^{2\pi}d\sigma d\rho
    \left[(D_{\tau}X^i)^2 -\frac{1}{2L^2}\{X^{i},X^{j}\}^2\right]
	\label{MLCgauge},\\
  &&D_{\tau}X^i=\ptau X^i -\frac{1}{L}\{A,X^i\},\\
  &&\{A,B\}\equiv \psig A\prho B -\prho A\psig B,
\end{eqnarray}
where $i,j=1,2,\cdots,9$, $T$ is the tension of the supermembrane and
$L$ is an arbitrary length parameter.\footnote{Note that the mass
dimensions of the world-volume parameters, $\tau,\sigma$ and $\rho$,
are $0$.}
It is easy to see that the parameter $L$ can be changed for $L'$ just
by a rescaling of $\tau \rightarrow \tau L/L'$.
Eq.(\ref{MLCgauge}) represents a gauge theory of the area-preserving
diffeomorphism.
In this paper, we consider three kinds of the eleven-dimensional
target space: $R^{11}$, $R^{10}\times S^1$ and $R^{9}\times T^2$.
On $R^{11}$, the target-space coordinates $X^i$ and gauge field
$A$ are expanded as\footnote{For simplicity, we consider only
toroidal supermembrane in this paper.}
\begin{eqnarray}
 X^i(\sigma,\rho)&=&\sum_{k_1,k_2 = -\infty}^{\infty}X^i_{(k_1,k_2)}\,
	e^{ik_1\sigma+ik_2\rho},\label{mode1}\\
 A(\sigma,\rho)&=&\sum_{k_1,k_2=-\infty}^{\infty}A_{(k_1,k_2)}\,
	e^{ik_1\sigma+ik_2\rho}.\label{mode2}
\end{eqnarray}

On $R^{10}\times S^1$, we can consider the wrapped supermembrane.
We take $X^9$ as a coordinate of $S^1$ with the radius $L_1$ .
Then $X^i\ (i=k,9\ \,(k=1,2,\cdots,8))$ and $A$
are expanded as
\begin{eqnarray}
 X^9(\sigma,\rho)&=&w_1L_1\rho +\sum_{k_1,k_2=-\infty}^{\infty}
	 Y^1_{(k_1,k_2)}\,e^{ik_1\sigma+ik_2\rho}
	\equiv w_1L_1\rho + Y^1(\sigma,\rho),\label{mode'1}\\
 X^k(\sigma,\rho)&=&\sum_{k_1,k_2 = -\infty}^{\infty}X^k_{(k_1,k_2)}\,
	e^{ik_1\sigma+ik_2\rho},\label{mode'2}\\
 A(\sigma,\rho)&=&\sum_{k_1,k_2=-\infty}^{\infty}A_{(k_1,k_2)}\,
	e^{ik_1\sigma+ik_2\rho},\label{mode'3}
\end{eqnarray}
where $w_1$ is an integer.
If we regard $X^9$ as the 11th direction, this supermembrane
corresponds to the light-cone type-IIA superstring on $R^{10}$ via the
double-dimensional reduction, which has a tension
$1/(2\pi\alpha') = 2\pi L_1 T$ \cite{DHIS}.\footnote{The
double-dimensional reduction was discussed classically in
Ref.\cite{DHIS}. In quantum mechanically, it is subtle whether such a
reduction is realized or not \cite{Rus,SY,UY}.}

Finally, we consider the wrapped supermembrane on $R^{9}\times T^2$.
We take $X^8$ and $X^9$ as the coordinates of two cycles of $T^2$.
Then we can expand $X^i\ (i=m,8,9\ \,(m=1,2,\cdots,7))$ and $A$ as
\begin{eqnarray}
 X^9(\sigma,\rho) &=&w_1L_1\rho +\sum_{k_1,k_2=-\infty}^{\infty}
	 Y^1_{(k_1,k_2)}\,e^{ik_1\sigma+ik_2\rho}
	=  w_1L_1\rho + Y^1(\sigma,\rho),\label{mode''1}\\
 X^8(\sigma,\rho)&=&w_2L_2\sigma +\sum_{k_1,k_2=-\infty}^{\infty}
	 Y^2_{(k_1,k_2)}\,e^{ik_1\sigma+ik_2\rho}
	 \equiv w_2L_2\sigma + Y^2(\sigma,\rho),\label{mode''2}\\
 X^m(\sigma,\rho)&=&\sum_{k_1,k_2 = -\infty}^{\infty}X^m_{(k_1,k_2)}\,
	e^{ik_1\sigma+ik_2\rho},\label{mode''3}\\
 A(\sigma,\rho)&=&\sum_{k_1,k_2=-\infty}^{\infty}A_{(k_1,k_2)}\,
	e^{ik_1\sigma+ik_2\rho},\label{mode''4}
\end{eqnarray}
where $L_1$ and $L_2$ are the radii of two cycles of $T^2$
and $w_1$ and $w_2$ are integers.
The wrapping number on $T^2$ is given by $w_1w_2$.
If we also consider $X^9$ as the 11th direction, this supermembrane
corresponds to the light-cone type-IIA superstring on $R^{9}\times
S^1$ (radius $L_2$), which is  T-dual to the light-cone type-IIB
superstring on $R^{9}\times S^1$ (radius $\tilde{L}_2=\alpha'/L_2$).

%%%%%%%%%%%%%%%%%%%%%%%%%%%%%%%%%%
\section{From supermembrane to 0+1-D super Yang-Mills}\label{S:0+1}
%%%%%%%%%%%%%%%%%%%%%%%%%%%%%%%%%%
In this section, we consider the light-cone supermembrane on $R^{11}$,
where the mode expansions are given in eqs.(\ref{mode1}) and
(\ref{mode2}).
It is known that we shall obtain 0+1-dimensional maximally
supersymmetric $U(N)$ Yang-Mills theory through the matrix
regularization of the supermembrane \cite{Hop,dWHN}. We pursue the
procedure as a warm-up before the subsequent sections.

The matrix regularization is the following mathematical procedure:
First, we introduce the noncommutativity for the space-sheet
coordinates of the supermembrane, $[\,\sigma,\rho\,]=i\mtheta$
($\mtheta$: constant).
This noncommutativity is encoded in the star product for functions on
the space sheet,
\begin{equation}
 f*g=f\exp\left(i\frac{1}{2}\mtheta\,\epsilon^{\alpha\beta}\,
    \overleftarrow{\p}_{\alpha}\overrightarrow{\p}_{\beta}\right)\,g.
	\qquad(\alpha,\beta=\sigma,\rho)
\end{equation}
Then, the star-commutator for the Fourier modes in eqs.(\ref{mode1})
and (\ref{mode2}) is given by \cite{FZ}
\begin{equation}
  \scom{e^{ik_1\sigma+ik_2\rho},e^{ik'_1\sigma+ik'_2\rho}}
   =-2i\sin\left(\frac{1}{2}\mtheta\,k\times k'\right)\,
	e^{i(k_1+k'_1)\sigma+i(k_2+k'_2)\rho}.\label{alg}
\end{equation}
In the $\mtheta\to0$ limit, the space-sheet Poisson bracket is
obtained,
\begin{equation}
  \{\,f,g\,\}=-i \lim_{\mtheta\to 0}\mtheta^{-1}\,\scom{f,g}\,.
\end{equation}
If $\mtheta=2\pi q/N$ where $q$ and $N$ are mutually prime integers,
the sine functions in the structure constants have zeros.
Henceforth we set $\mtheta=2\pi /N$ with $N=2M+1$ being odd number
for simple presentation here.
Then, the Fourier modes $e^{ipN\sigma}$, $e^{irN\rho}$
($p,r\in\mathbb{Z}$) commute with any modes and hence they are central
elements in the star-commutator algebra.
Thus we can identify them with the identity operator
consistently in the algebra and then we obtain the following
equivalences,\footnote{In eqs.(\ref{ide1}) and (\ref{ide2}), the sign
factors depending on the Fourier modes are attached. These factors
can be eliminated by choosing $\mtheta=4\pi /N$ (for odd
$N$). Actually, in our previous paper \cite{UY3}, such a convention was
adopted. In a mathematical point of view, this is nothing but a
convention. In this paper, however, we adopt $\mtheta=2\pi /N$ and we
will explain the physical meaning later.}
\begin{eqnarray}
 e^{i(k_1+pN)\sigma+ik_2\rho}&\approx&
	(-1)^{pk_2} e^{ik_1\sigma+ik_2\rho},\label{ide1}\\
 e^{ik_1\sigma+i(k_2+rN)\rho}&\approx&
	 (-1)^{rk_1}e^{ik_1\sigma+ik_2\rho}.\label{ide2}
\end{eqnarray}
Under the equivalences, the infinite dimensional algebra is
consistently truncated to the finite dimensional algebra
$\mathfrak{u}(N)$.
Then, the number of the modes $e^{ik_1\sigma+ik_2\rho}$ are
restricted to $k_1,k_2=0,\pm1,\pm2,\cdots,\pm M$ and the truncated
mode expansions are given by
\begin{eqnarray}
 X^i(\sigma,\rho)&=&\sum_{k_1,k_2 = -M}^{M}X^i_{(k_1,k_2)}\,
	e^{ik_1\sigma+ik_2\rho},\label{mode1'}\\
 A(\sigma,\rho)&=&\sum_{k_1,k_2=-M}^{M}A_{(k_1,k_2)}\,
	e^{ik_1\sigma+ik_2\rho}.\label{mode2'}
\end{eqnarray}

Next, we give a matrix representation for the $N^2$ generators
of $\mathfrak{u}(N)$, $\{e^{ik_1\sigma+ik_2\rho}\,|\, k_1,k_2=0,\pm1,
\pm2, \cdots, \pm M\}$.
Actually, the generators are represented by $N\times N$ matrices
\cite{FZ}
\begin{eqnarray}
  e^{ik_1\sigma+ik_2\rho}\to\lambda^{-k_1k_2/2}\,V^{k_2}\,U^{k_1},
\label{matrix}
\end{eqnarray}
where $\lambda \equiv e^{i2\pi/N}$,
$U$ and $V$ are the clock and shift matrices, respectively,
\begin{eqnarray}
  U &=& \left(\begin{array}{ccccc}
	1 & & & &  \lower10pt\hbox{\Large 0}\\[-5pt]
	 & \lambda& & &  \\
	 &  & \lambda^2& &  \\
	 &  &  & \ddots &\\
	&\hbox{\Large0} & & &\lambda^{N-1}\end{array}\right),
	\label{eq:U}\\
 V &=& \left(\begin{array}{ccc@{}c@{}cc}
	0& 1&  & & & \\
	 & 0& 1& & & \\
	\vdots &  &  &\ddots&\ddots  & \\
	0&  &  & & 0& 1\\
	1& 0&  &\cdots & &0\end{array}\right) .
\end{eqnarray}
$U$ and $V$ have the following properties,
\begin{eqnarray}
	U^N&=&V^N=1,\label{pro1}\\
	VU&=&\lambda UV.\label{pro2}
\end{eqnarray}
For $k<0$,
$U^k\equiv(U^{\dagger})^{-k},\,V^k\equiv(V^{\dagger})^{-k}$.
By using eq.(\ref{pro2}), it is easy to show that the representation
(\ref{matrix}) satisfy the commutator (\ref{alg}) with
$\mtheta=2\pi/N$.
Furthermore, from eq.(\ref{pro1}), equivalences (\ref{ide1}) and
(\ref{ide2}) are satisfied.
{}From eqs.(\ref{matrix}) and (\ref{eq:U}), we see that  $N\times N$
matrices of the zero-modes with respect to $\rho$ ($k_2=0$ in
(\ref{matrix})) are diagonalized. However, this basis is a convention.
We can transform to another basis where the zero-modes with respect to
$\sigma$ ($k_1=0$ in (\ref{matrix})) are diagonalized by using the
following unitary matrix,
\begin{eqnarray}
 S&=& \frac{1}{\sqrt{N}}
    \renewcommand{\arraystretch}{1.5}\left(\begin{array}{ccccc}
	1& 1& 1&  & 1\\
	1& \lambda& \lambda^2& \cdots &\lambda^{(N-1)}\\
	1 & \lambda^2 & \lambda^4& \cdots &\lambda^{2(N-1)} \\
	\vdots &\vdots  &\vdots &  \ddots &\vdots\\
	1&\lambda^{N-1}&\lambda^{2(N-1)}&\cdots&\lambda^{(N-1)^2}
    \end{array}\right)\,,
\end{eqnarray}
since  $U$ and $V$ are transformed
as $S^{\dagger}VS=U,\,S^{\dagger}US=V^{-1}$.
According to (\ref{matrix}), the truncated mode
expansions (\ref{mode1'}) and (\ref{mode2'}) are represented
by $N\times N$ matrices,
\begin{eqnarray}
 X^i(\sigma,\rho)&\to&
 X^i=\sum_{k_1,k_2 = -M}^{M}X^i_{(k_1,k_2)}\,
	\lambda^{-k_1k_2/2}\,V^{k_2}\,U^{k_1},\label{mode1''}\\
 A(\sigma,\rho)&\to&
 A=\sum_{k_1,k_2 = -M}^{M}A_{(k_1,k_2)}\,
	\lambda^{-k_1k_2/2}\,V^{k_2}\,U^{k_1}.\label{mode2''}
\end{eqnarray}

Finally, we show that the matrix-regularized action of the light-cone
supermembrane on $R^{11}$ agrees with that of 0+1-dimensional
maximally supersymmetric $U(N)$ Yang-Mills theory.
In the matrix regularization, the functions $X^i,A$ of $\sigma$ and
$\rho$ are represented by the $N\times N$ matrices (\ref{mode1''}) and
(\ref{mode2''}) and the Poisson bracket and the double integral are
represented as follows,
\begin{eqnarray}
 \{\,\cdot\,,\,\cdot\,\} &\to&
	-i\frac{N}{2\pi}[\,\cdot\,,\,\cdot\,],\\
 \int_0^{2\pi}d\sigma d\rho &\to& \frac{(2\pi)^2}{N}\,\mbox{Tr}.
\end{eqnarray}
{}From these results and a rescaling $\tau\to \tau/N$, the action
(\ref{MLCgauge}) is mapped to
\begin{eqnarray}
 S_{0+1} &=&(2\pi)^2 LT\int d\tau
   \mbox{Tr}\left[ (D_{\tau}X^i)^2
    +\frac{1}{2(2\pi L)^2}[X^{i},X^{j}]^2\right]\!,\\
 &&D_{\tau}X^i=\ptau X^i +i\frac{1}{2\pi L}[A,X^i],
\end{eqnarray}
which is just a bosonic part of 0+1-dimensional maximally
supersymmetric $U(N)$ Yang-Mills theory, i.e., Matrix theory.

%%%%%%%%%%%%%%%%%%%%%%%%%%%%%%%%%%
\section{From supermembrane to 1+1-D super Yang-Mills}\label{S:1+1}
%%%%%%%%%%%%%%%%%%%%%%%%%%%%%%%%%%
In this section, we consider the light-cone wrapped supermembrane on
$R^{10}\times S^1$, which has the mode expansions
(\ref{mode'1})-(\ref{mode'3}).
{}From this supermembrane, we can obtain 1+1-dimensional maximally
supersymmetric $U(N)$ Yang-Mills theory by introducing the
noncommutativity $[\sigma,\rho]=i2\pi/N$ ($N=2M+1$ : odd number) and
giving a matrix representation of the star-commutator algebra.
We shall immediately notice that we need to handle the wrapping of the
supermembrane this time. It was pointed out that we should add a
linear term $\rho$ to the generators of the star-commutator algebra
\cite{Ced}. Then the star commutators are given by
\begin{eqnarray}
  \scom{e^{ik_1\sigma+ik_2\rho},e^{ik'_1\sigma+ik'_2\rho}}
   &=&-2i\sin\left(\frac{\pi}{N}\,k\times k'\right)\,
	e^{i(k_1+k'_1)\sigma+i(k_2+k'_2)\rho},
\label{alg1}\\
\scom{\rho,e^{ik_1\sigma+ik_2\rho}}&=&
	\frac{2\pi k_1}{N}\, e^{ik_1\sigma+ik_2\rho}.\label{alg2}
\end{eqnarray}
In this algebra, we should notice that the Fourier modes
$e^{ipN\sigma}$ cannot be the central elements and hence the
equivalence (\ref{ide1}) is no longer valid this time.
On the other hand,  $e^{irN\rho}$ are the central elements and the
equivalence (\ref{ide2}) is still valid.
Then, we can consistently truncate only the Fourier modes with respect
to $\rho$ and the truncated generators are given by
$\{e^{ik_1\sigma+ik_2\rho},\rho\,|\,k_1=0,\pm1,\pm2,\cdots,\pm\infty,
\,k_2=0,\pm1,\pm2,\cdots,\pm M\}$ \cite{Ced}.\footnote{Although we
cannot identify $e^{ipN\sigma}$ with the central elements, they form
an ideal of the truncated star-commutator algebra. Hence we can
consider the quotient by this ideal and the quotient is affine
$\mathfrak{su}(N)$ except for a central element \cite{Ced,UY3}.}
The truncated mode expansions are given by
\begin{eqnarray}
 X^9(\sigma,\rho)
    &=&w_1L_1\rho +\sum_{k_1=-\infty}^{\infty}\sum_{k_2=-M}^{M}
	 Y^1_{(k_1,k_2)}\,e^{ik_1\sigma+ik_2\rho}\nn
    &=&w_1L_1\rho +\sum_{p=-\infty}^{\infty}\sum_{q=-M}^{M}
	\sum_{k=-M}^{M} Y^1_{(pN+q,k)}\,e^{i(pN+q)\sigma+ik\rho},
	\label{mode'1'}\\
 X^k(\sigma,\rho)&=& \sum_{k_1= -\infty}^{\infty}
	\sum_{k_2= -M}^{M}X^k_{(k_1,k_2)}\,e^{ik_1\sigma+ik_2\rho}\nn
 &=&\sum_{p=-\infty}^{\infty}\sum_{q=-M}^{M}\sum_{k=-M}^{M}
    X^k_{(pN+q,k)}\,e^{i(pN+q)\sigma+ik\rho},\label{mode'2'}\\
 A(\sigma,\rho)&=&\sum_{k_1= -\infty}^{\infty}
	\sum_{k_2= -M}^{M}A_{(k_1,k_2)}\,e^{ik_1\sigma+ik_2\rho}\nn
  &=&\sum_{p=-\infty}^{\infty}\sum_{q=-M}^{M}\sum_{k=-M}^{M}
	A_{(pN+q,k)}\,e^{i(pN+q)\sigma+ik\rho}.\label{mode'3'}
\end{eqnarray}

Now we give a matrix representation of generators
$\{e^{i(pN+q)\sigma+ik\rho},\rho\,|\,p=0,\pm1,\pm2,\cdots,\pm\infty,
$ $q,k=0,\pm1,\pm2,\cdots,\pm M\}$.
Actually, the generators are represented as $N\times N$ matrices
with a continuous parameter $\theta_1$ \cite{UY3},
\begin{eqnarray}
  e^{i(pN+q)\sigma+ik\rho} &\rightarrow& e^{i(pN+q)\theta_1/N}
    \lambda^{-kq/2}\,V^{k}\,U^{q},\label{matrix1}\\
  \rho&\rightarrow&-2\pi i \p_{\theta_1} I,\label{matrix2}
\end{eqnarray}
where $I$ is the $N\times N$ unit matrix.
{}From eqs.(\ref{pro1}) and (\ref{pro2}), it is easy to see that the
representation (\ref{matrix1}) and (\ref{matrix2}) satisfy the
commutators (\ref{alg1}) and (\ref{alg2}) and also the equivalence
(\ref{ide2}).
In the $N\times N$ matrix representation (\ref{matrix1}) and
(\ref{matrix2}), the truncated mode expansions
(\ref{mode'1'})-(\ref{mode'3'}) are given by
\begin{eqnarray}
 X^9(\sigma,\rho) &\to& -2\pi i w_1L_1\p_{\theta_1} I
	+Y^1(\theta_1)\nn
  &=&-2\pi i w_1L_1\p_{\theta_1} I+ \sum_{p=-\infty}^{\infty}
	\sum_{q=-M}^{M}\sum_{k=-M}^{M} Y^1_{(pN+q,k)}
	e^{i(pN+q)\theta_1/N}\lambda^{-kq/2}V^{k}U^{q},\label{mode'1''}\\
 X^k(\sigma,\rho)&\to&
  X^k(\theta_1)=\sum_{p=-\infty}^{\infty}\sum_{q=-M}^{M}\sum_{k=-M}^{M}
    X^k_{(pN+q,k)}\,e^{i(pN+q)\theta_1/N}
	\lambda^{-kq/2}V^{k}U^{q},\label{mode'2''}\\
 A(\sigma,\rho) &\to&A(\theta_1)=\sum_{p=-\infty}^{\infty}
	\sum_{q=-M}^{M}\sum_{k=-M}^{M} A_{(pN+q,k)}\,
	e^{i(pN+q)\theta_1/N}\lambda^{-kq/2}V^{k}U^{q}.\label{mode'3''}
\end{eqnarray}
Here we find that $Y^1(\theta_1), X^k(\theta_1)$ and $A(\theta_1)$
satisfy the boundary conditions,
\begin{eqnarray}
  Y^1(\theta_1+2\pi)&=&V Y^1(\theta_1)V^{\dagger},\label{bc1a}\\
  X^k(\theta_1+2\pi)&=&V X^k(\theta_1)V^{\dagger},\\
  A(\theta_1+2\pi)&=&V A(\theta_1)V^{\dagger},\label{bc3a}
\end{eqnarray}
because of $VUV^{\dagger}=\lambda U$.
This boundary conditions means that via the double-dimensional
reduction, the wrapped supermembrane becomes to correspond to a long
string in matrix string theory.
In Ref.\cite{SY}, the boundary conditions were assumed, while they are
deducible in our case \cite{UY3}.

Next, we show that the matrix-regularized action of the light-cone
supermembrane on $R^{10}\times S^1$ agrees with that of
1+1-dimensional maximally supersymmetric $U(N)$ Yang-Mills theory.
In such a truncation, the functions $X^9,X^k,A$ of $\sigma$ and $\rho$
are represented by the matrices (\ref{mode'1''})-(\ref{mode'3''}) and
the Poisson bracket and the double integral are represented as follows,
\begin{eqnarray}
 \{\,\cdot\,,\,\cdot\,\} &\to&
	-i\frac{N}{2\pi}[\,\cdot\,,\,\cdot\,]~,\\
 \int_0^{2\pi}d\sigma d\rho &\to& \frac{2\pi}{N}\int_0^{2\pi}
	d\theta_1\,\mbox{Tr}~.
\end{eqnarray}
{}From these results and a rescaling $\tau\to\tau/N$, the action
(\ref{MLCgauge}) in the $w_1=1$ case \footnote{Henceforth we set
$w_1=1$ for simplicity. See Ref.\cite{UY3} for the discussion in the
case of an arbitrary integer $w_1$.} is mapped to \cite{SY,Ced,UY3}
\begin{eqnarray}
 S_{1+1} &=&\frac{2\pi LT}{2}\int d\tau
	\int_0^{2\pi}d\theta_1\,\mbox{Tr}\Biggl[(F_{\tau\theta_1})^2
	+ (D_{\tau}X^k)^2\nn
    &&\hspace{25ex} -(D_{\theta_1}X^k)^2
	+\frac{1}{2(2\pi L)^2}\,[X^{k},X^{l}]^2\Biggr],\label{MS}\\
  &&F_{\tau\theta_1}=\ptau Y^1 -\frac{L_1}{L}\p_{\theta_1}A
	+i\frac{1}{2\pi L}\,[A,Y^1],\label{MS2}\\
  &&D_{\tau}X^k=\ptau X^k +i\frac{1}{2\pi L}\,[A,X^k],\label{MS3}\\
  &&D_{\theta_1}X^k=\frac{L_1}{L} \p_{\theta_1} X^k
	+i\frac{1}{2\pi L}\,[Y^1,X^k].\label{MS4}
\end{eqnarray}
Here we should note that the fields $Y^1(\theta_1), X^k(\theta_1)$,
$A(\theta_1)$ have mass dimension $-1$ and the parameters $\tau,
\theta_1$ have mass dimension $0$.
We rewrite the action (\ref{MS}) to the standard form of Yang-Mills
theory.
The kinetic term of the gauge field in $D$ dimensions is given by
\begin{eqnarray}
 S_{YM}=-\frac{1}{4g_{YM}^2}\int d^D x
	\mbox{Tr} F_{\mu\nu}F^{\mu\nu},\label{sta1}\\
 F_{\mu\nu}=\p_{\mu}A_{\nu}-\p_{\nu}A_{\mu}
	+i[A_{\mu},A_{\nu}],\label{sta2}
\end{eqnarray}
where the mass dimensions of the gauge field $A_{\mu}(x)$, the
parameter $x^{\mu}$ and the gauge coupling constant $g_{YM}$
are $1$, $-1$ and $2-D/2$, respectively.
In order to adjust their mass dimensions we shall introduce some
dimensionful constants $\alpha, \Sigma, \Sigma_1$ for the time being
and make a change of variables,
\begin{eqnarray}
  Y^1(\theta_1)&\to&\alpha A_1(x^1)\,,\\
  X^k(\theta_1)&\to&\alpha \phi^k(x^1)\,,\\
  A(\theta_1)&\to&\alpha A_0(x^1)\,,\\
  \theta_1 &\to& \frac{x^1}{\Sigma_1}\,,\\
  \tau &\to& \frac{x^0}{\Sigma}\,,
\end{eqnarray}
where $\alpha$ has mass dimension $-2$ and $\Sigma$ and $\Sigma_1$
have mass dimension $-1$.
Then, the action (\ref{MS}) is rewritten by
\begin{eqnarray}
 S_{1+1} &=&\frac{2\pi LT}{2}\frac{1}{\Sigma_1\Sigma}
	\int dx^0 \int_0^{2\pi\Sigma_1}\!\!dx^1\,
	\mbox{Tr}\Biggl[ (F_{\tau\theta_1})^2+(D_{\tau}X^k)^2\nn
  &&\hspace{30ex}-(D_{\theta_1}X^k)^2
	+\frac{\alpha^4}{2(2\pi L)^2}\,
	[\phi^{k},\phi^{l}]^2\Biggr],\label{MS'}\\
 &&F_{\tau\theta_1}=\Sigma\alpha\p_{0}A_1 -
	\frac{L_1}{L}\,\Sigma_1\alpha\p_{1}A_0
	+i\frac{\alpha^2}{2\pi L}\,[A_0,A_1],\label{MS2'}\\
  &&D_{\tau}X^k=\Sigma\alpha\p_{0} \phi^k +i\frac{\alpha^2}{2\pi L}
	\,[A_0,\phi^k],\label{MS3'}\\
 &&D_{\theta_1}X^k=\frac{L_1}{L}\Sigma_1\alpha\p_{1}\phi^k
	 +i\frac{\alpha^2}{2\pi L}\,[A_1,\phi^k].\label{MS4'}
\end{eqnarray}
In order that the field strength (\ref{MS2'}) takes the standard
form (\ref{sta2}), the constants should satisfy the following
relations,
\begin{eqnarray}
  \Sigma&=&\frac{\alpha}{2\pi L},\\
  \Sigma_1&=&\frac{\alpha}{2\pi L_1}. \label{T-dual}
\end{eqnarray}
Notice that {\it eq.(\ref{T-dual}) represents the T-duality which
relates the length $\Sigma_1$ in the super Yang-Mills theory and the
length $L_1$ in M-theory.}
{}From the D-brane viewpoint \cite{Sei,Sen}, this super Yang-Mills
theory is regarded as the low energy effective theory of $N$ D1-branes
and hence this is exactly the T-duality between
D0-branes and D1-branes \cite{Tay}.
However, we should notice that we have considered from the membrane
viewpoint in this paper and hence we have given another derivation of
the T-duality of eq.(\ref{T-dual}).\footnote{From the D-brane
viewpoint based on Refs.\cite{Sei,Sen}, where the 11th direction is
chosen as the longitudinal direction, the parameter $\alpha$ should be
identified with the inverse of the string tension $2\pi RT$ where $R$
is a radius of the $x^-$ direction in DLCQ \cite{GRT,FHRS}. However,
in this paper we have taken a different approach and hence $\alpha$ is
an arbitrary parameter having the mass dimension $-2$ in our case.}
Ultimately, we have obtained the standard form of a bosonic part of
1+1-dimensional maximally supersymmetric $U(N)$ Yang-Mills theory,
i.e., matrix string theory,
\begin{eqnarray}
S_{1+1} &=&\frac{1}{2g_{YM}^2} \int dx^0
	\int_0^{2\pi\Sigma_1}\!\!dx^1\,
	 \mbox{Tr}\left[ (F_{01})^2+(D_{0}\phi^k)^2 -(D_{1}\phi^k)^2
	+\frac{1}{2}[\phi^{k},\phi^{l}]^2\right],\label{MS''}\\
 &&F_{01}=\p_0 A_1 - \p_{1}A_0 +i[A_0,A_1],\label{MS2''}\\
 &&D_{0}\phi^k=\p_0 \phi^k +i[A_0,\phi^k],\label{MS3''}\\
 &&D_{1}\phi^k=\p_{1} \phi^k +i[A_1,\phi^k],\label{MS4''}
\end{eqnarray}
with the boundary conditions
\begin{eqnarray}
  A_0(x^1+2\pi\Sigma_1)&=&V A_0(x^1)V^{\dagger},\\
  A_1(x^1+2\pi\Sigma_1)&=&V A_1(x^1)V^{\dagger},\\
 \phi^k(x^1+2\pi\Sigma_1)&=&V \phi^k(x^1)V^{\dagger},
\end{eqnarray}
where $g_{YM}$ is the gauge coupling constant of mass dimension 1,
which is given by
$g_{YM}^2= (2\pi)^{-3} \Sigma_1^{-2}L_1^{-3}T^{-1}$.
We also define the dimensionless gauge coupling constant
$\tilde{g}_{YM}$ by
\begin{eqnarray}
 \tilde{g}_{YM}^2\equiv g_{YM}^2(2\pi \Sigma_1)^2
	=(2\pi)^{-1}L_1^{-3}T^{-1}
	= 2\pi\,\frac{l_{11}^3}{L_1^3},\label{YMconst}
\end{eqnarray}
where eleven-dimensional Planck length $l_{11}$ is defined
by $T\equiv (2\pi)^{-2}l_{11}^{-3}$.
The dimensionless gauge coupling constant in eq.(\ref{YMconst})
agrees with that in Ref.\cite{FHRS} including the numerical
constant.\footnote{Note that the parameters $\Sigma$ and $L$ in
Ref.\cite{FHRS} represent the circumferences but not the radii.}
Note that in Ref.\cite{FHRS} such a coupling constant was
derived by regarding the super Yang-Mills theory as the low energy
effective theory of D-branes (D-brane viewpoint), while we have
obtained the coupling constant by the matrix regularization of the
light-cone supermembrane in this paper (membrane viewpoint).
Here we make a comment on the convention of the noncommutative
parameter $\mtheta$. It is nothing but a mathematical convention
whether we adopt $\mtheta=2\pi/N$, $\mtheta=4\pi/N$ and so on in
the matrix regularization. However, if we would not adopt $\mtheta=2\pi
/N$, eq.(\ref{YMconst}) has a different numerical constant which does
not agree with that in Ref.\cite{FHRS}.
Thus the $N$ in a choice of $\mtheta=2\pi /N$ has the physical meaning
of the $N$ in the light-cone momentum $p^+=N/R$ in DLCQ.
 
Finally, we consider $X^9$ as the 11th direction and relate to
type-IIA superstring theory.
The precise relation is that the string tension $(2\pi\alpha')^{-1}$
and string coupling constant $g_s$ are given by \cite{DHIS,Wit}
\begin{eqnarray}
  \frac{1}{2\pi \alpha'}&=&2\pi L_1T,\\
  g_s&=&\frac{L_1}{\sqrt{\alpha'}}\,.
\end{eqnarray}
Evidently, the dimensionless gauge coupling constant $\tilde{g}_{YM}$
and string coupling constant $g_s$ are inversely related to one
another \cite{FHRS}
\begin{equation}
	\tilde{g}_{YM}^2=\frac{2\pi}{g_s^2}\,.
\end{equation}
Thus the super Yang-Mills theory in the strong coupling limit is
equivalent to free type-IIA superstring theory in the light-cone gauge.

%%%%%%%%%%%%%%%%%%%%%%%%%%%%%%%%%%
\section{From supermembrane to 2+1-D super Yang-Mills}\label{S:2+1}
%%%%%%%%%%%%%%%%%%%%%%%%%%%%%%%%%%
In this section, we consider the light-cone wrapped supermembrane on
$R^{9}\times T^2$, which has the mode expansions
(\ref{mode''1})-(\ref{mode''4}).
Starting from this supermembrane, we can obtain 2+1-dimensional
maximally supersymmetric $U(N)$ Yang-Mills theory by introducing the
noncommutativity $[\sigma,\rho]=i2\pi/N$ ($N=2M+1$ : odd number) and
giving a matrix representation of the star-commutator algebra.
In this case, we need to add two linear terms, $\sigma$ and $\rho$, to
the generators of the star-commutator algebra.
Then the star commutators are given by
\begin{eqnarray}
  \scom{e^{ik_1\sigma+ik_2\rho},e^{ik'_1\sigma+ik'_2\rho}}
  &=&-2i\sin\left(\frac{\pi}{N}\,k\times k'\right)\,
	e^{i(k_1+k'_1)\sigma+i(k_2+k'_2)\rho}\,,\label{alg'1}\\
  \scom{\sigma,e^{ik_1\sigma+ik_2\rho}}&=&
    -\frac{2\pi k_2}{N}\, e^{ik_1\sigma+ik_2\rho}\,,\label{alg'2}\\
  \scom{\rho,e^{ik_1\sigma+ik_2\rho}}&=&
    \frac{2\pi k_1}{N}\, e^{ik_1\sigma+ik_2\rho}\,,\label{alg'3}\\
  \scom{\sigma,\rho}&=&i\frac{2\pi}{N}\,.\label{alg'4}
\end{eqnarray}
In this algebra, both $e^{ipN\sigma}$ and $e^{irN\rho}$ cannot be the
central elements and hence the equivalences (\ref{ide1}) and
(\ref{ide2}) do not hold.
Thus in this case, we cannot consistently truncate the Fourier modes
at all,
\begin{eqnarray}
  X^9(\sigma,\rho)&=& w_1L_1\rho +\sum_{k_1,k_2=-\infty}^{\infty}
	 Y^1_{(k_1,k_2)}\,e^{ik_1\sigma+ik_2\rho}\nn
  &=& w_1L_1\rho +\sum_{p,r=-\infty}^{\infty}
	\sum_{q,s=-M}^{M}Y^1_{(pN+q,rN+s)}
	\,e^{i(pN+q)\sigma+i(rN+s)\rho}, \label{mode'''1}\\
 X^8(\sigma,\rho)&=&w_2L_2\sigma +\sum_{k_1,k_2=-\infty}^{\infty}
	 Y^2_{(k_1,k_2)}\,e^{ik_1\sigma+ik_2\rho}\nn
    &=&w_2L_2\sigma +\sum_{p,r=-\infty}^{\infty}
	\sum_{q,s=-M}^{M} Y^2_{(pN+q,rN+s)}
	\,e^{i(pN+q)\sigma+i(rN+s)\rho},\label{mode'''2}\\
 X^m(\sigma,\rho)&=&\sum_{k_1,k_2 = -\infty}^{\infty}X^m_{(k_1,k_2)}\,
	e^{ik_1\sigma+ik_2\rho}\nn
 &=&\sum_{p,r = -\infty}^{\infty}\sum_{q,s=-M}^{M}X^m_{(pN+q,rN+s)}\,
	e^{i(pN+q)\sigma+i(rN+s)\rho}\label{mode'''3}\\
 A(\sigma,\rho)&=&\sum_{k_1,k_2=-\infty}^{\infty}A_{(k_1,k_2)}\,
	e^{ik_1\sigma+ik_2\rho}\nn
 &=&\sum_{p,r = -\infty}^{\infty}\sum_{q,s=-M}^{M}A_{(pN+q,rN+s)}\,
	e^{i(pN+q)\sigma+i(rN+s)\rho}.\label{mode'''4}
\end{eqnarray}

Although the consistent truncation does not exist, we can give a
matrix representation of the generators
$\{e^{i(pN+q)\sigma+i(rN+s)\rho},\sigma,\rho\,|\,p,r=0,
\pm1,\pm2,\cdots,\pm\infty,\ q,s=0,\pm1,\pm2,\cdots,\pm M\}$.
Actually, the generators are represented as $N\times N$ matrices
with two continuous parameters $\theta_1,\theta_2$,
\begin{eqnarray}
 e^{i(pN+q)\sigma+i(rN+s)\rho} &\to& e^{i(pN+q)\theta_1/N}
   e^{-i(rN+s)\theta_2/N}\lambda^{-sq/2}V^{s}U^{q},\label{matrix1'}\\
 \rho&\to&-2\pi i \p_{\theta_1} I,\label{matrix2'}\\
 \sigma&\to&-2\pi i \p_{\theta_2}I
	+\frac{\theta_1}{N}I.\label{matrix3'}
\end{eqnarray}
Due to the properties of $U$ and $V$ in eqs.(\ref{pro1}) and
(\ref{pro2}), it is easy to show that representation
(\ref{matrix1'})-(\ref{matrix3'}) satisfy the commutators
(\ref{alg'1})-(\ref{alg'4}).
Then in the $N\times N$ matrix representation
(\ref{matrix1'})-(\ref{matrix3'}), mode expansions
(\ref{mode'''1})-(\ref{mode'''4}) are given by
\begin{eqnarray}
 X^9(\sigma,\rho)&\to& -2\pi iw_1L_1\p_{\theta_1}I
	+Y^1(\theta_1,\theta_2)\nn
 &=& -2\pi iw_1L_1\p_{\theta_1}I\nn
 &&+\sum_{p,r=-\infty}^{\infty}\sum_{q,s=-M}^{M} Y^1_{(pN+q,rN+s)}
	e^{i(pN+q)\theta_1/N}e^{-i(rN+s)\theta_2/N}
	\lambda^{-sq/2}V^{s}U^{q}\,,\label{mode31m}\\
 X^8(\sigma,\rho)&\to& -2\pi iw_2L_2\p_{\theta_2}I
	+\frac{w_2L_2}{N}\theta_1 I +Y^2(\theta_1,\theta_2)\nn
 &=&-2\pi iw_2L_2\p_{\theta_2}I+\frac{w_2L_2}{N}\theta_1I\nn
 &&+\sum_{p,r=-\infty}^{\infty}\sum_{q,s=-M}^{M}
	 Y^2_{(pN+q,rN+s)}e^{i(pN+q)\theta_1/N}
	e^{-i(rN+s)\theta_2/N}\lambda^{-sq/2}V^{s}U^{q},\label{mode32m}\\
 X^m(\sigma,\rho)&\to&X^m(\theta_1,\theta_2)\nn
  &=&\sum_{p,r=-\infty}^{\infty}\sum_{q,s=-M}^{M}
	 X^m_{(pN+q,rN+s)}e^{i(pN+q)\theta_1/N}
	e^{-i(rN+s)\theta_2/N}\lambda^{-sq/2}V^{s}U^{q},\label{mode33m}\\
 A(\sigma,\rho)&\to&A(\theta_1,\theta_2)\nn
  &=&\sum_{p,r=-\infty}^{\infty}\sum_{q,s=-M}^{M} A_{(pN+q,rN+s)}
	e^{i(pN+q)\theta_1/N}
	e^{-i(rN+s)\theta_2/N}\lambda^{-sq/2}V^{s}U^{q}.\label{mode34m}
\end{eqnarray}
Here we find that $Y^1(\theta_1,\theta_2),Y^2(\theta_1,\theta_2),
X^m(\theta_1,\theta_2)$ and $A(\theta_1,\theta_2)$
satisfy the boundary conditions
\begin{eqnarray}
 Y^1(\theta_1+2\pi,\theta_2)&=&VY^1(\theta_1,\theta_2)V^{\dagger},\\
 Y^2(\theta_1+2\pi,\theta_2)&=&VY^2(\theta_1,\theta_2)V^{\dagger},\\
 X^m(\theta_1+2\pi,\theta_2)&=&VX^m(\theta_1,\theta_2)V^{\dagger},\\
 A(\theta_1+2\pi,\theta_2)&=&VA(\theta_1,\theta_2)V^{\dagger},\\
 Y^1(\theta_1,\theta_2+2\pi)&=&UY^1(\theta_1,\theta_2)U^{\dagger},\\
 Y^2(\theta_1,\theta_2+2\pi)&=&UY^2(\theta_1,\theta_2)U^{\dagger},\\
 X^m(\theta_1,\theta_2+2\pi)&=&UX^m(\theta_1,\theta_2)U^{\dagger},\\
 A(\theta_1,\theta_2+2\pi)&=&UA(\theta_1,\theta_2)U^{\dagger},
\end{eqnarray}
because of $VUV^{\dagger}=\lambda U$ and
$UVU^{\dagger}=\lambda^{-1}V$.

Next, we show that after the introduction of the noncommutativity
$[\sigma,\rho]=i2\pi/N$, the action of the light-cone wrapped
supermembrane on $R^{9}\times T^2$ agrees with that of 2+1-dimensional
maximally supersymmetric $U(N)$ Yang-Mills theory with constant
magnetic flux.
This is an example of the mapping from noncommutative gauge theory to
ordinary (commutative) gauge theory with constant magnetic flux
\cite{SW}.
Actually, the functions $X^9,X^8,X^m,A$ of $\sigma$ and $\rho$
are represented by the matrices (\ref{mode31m})-(\ref{mode34m}) and
the Poisson bracket and the double integral are represented as follows,
\begin{eqnarray}
 \{\,\cdot\,,\,\cdot\,\} &\to&
	-i\frac{N}{2\pi}[\,\cdot\,,\,\cdot\,],\\
 \int_0^{2\pi}d\sigma d\rho &\to& \frac{1}{N}\int_0^{2\pi}
	d\theta_1d\theta_2\,\mbox{Tr}.
\end{eqnarray}
{}From these results and a rescaling $\tau\to \tau/N$, the action
(\ref{MLCgauge}) in the case of $w_1=w_2=1$ \footnote{Henceforth,
we set $w_1=w_2=1$ for simplicity.}
is mapped to
\begin{eqnarray}
 S_{2+1}&=&\frac{LT}{2}\int d\tau \int_0^{2\pi}d\theta_1d\theta_2\,
    \mbox{Tr}\Biggl[ (F_{\tau\theta_1})^2+(F_{\tau\theta_2})^2
    -(F_{\theta_1\theta_2})^2+(D_{\tau}X^m)^2 -(D_{\theta_1}X^m)^2\nn
 &&\hspace{25ex}-(D_{\theta_2}X^m)^2
    +\frac{1}{2(2\pi L)^2}\,[X^{m},X^{n}]^2\Biggr],\label{D2}\\
 &&F_{\tau\theta_1}=\ptau Y^1 -\frac{L_1}{L}\,\p_{\theta_1}A
    +i\frac{1}{2\pi L}\,[A,Y^1],\\
 &&F_{\tau\theta_2}=\ptau Y^2 -\frac{L_2}{L}\,\p_{\theta_2}A
    +i\frac{1}{2\pi L}\,[A,Y^2],\\
 &&F_{\theta_1\theta_2}=\frac{1}{NL}L_1L_2 I
    +\frac{L_1}{L}\,\p_{\theta_1} Y^2 -\frac{L_2}{L}\,\p_{\theta_2}Y^1
    +i\frac{1}{2\pi L}\,[Y^1,Y^2],\\
 &&D_{\tau}X^m=\ptau X^k +i\frac{1}{2\pi L}\,[A,X^m],\\
 &&D_{\theta_1}X^m=\frac{L_1}{L}\,\p_{\theta_1} X^m
    +i\frac{1}{2\pi L}\,[Y^1,X^m],\\
 &&D_{\theta_2}X^m=\frac{L_2}{L}\,\p_{\theta_2} X^m
    +i\frac{1}{2\pi L}\,[Y^2,X^m].
\end{eqnarray}
Here we should note that the fields $Y^1(\theta_1,\theta_2),
Y^2(\theta_1,\theta_2), X^m(\theta_1,\theta_2), A(\theta_1,\theta_2)$
have mass dimension $-1$ and the parameters $\tau, \theta_1, \theta_2$
have mass dimension $0$.
We also rewrite the action (\ref{D2}) to the standard form of
Yang-Mills theory. In order to adjust the mass dimensions of the
fields and the parameters, we rewrite them by introducing some
dimensionful constants,
\begin{eqnarray}
 Y^1(\theta_1,\theta_2)&\to&\alpha A_1(x^1,x^2)\,,\\
 Y^2(\theta_1,\theta_2)&\to&\alpha A_2(x^1,x^2)\,,\\
 X^m(\theta_1,\theta_2)&\to&\alpha \phi^m(x^1,x^2)\,,\\
 A(\theta_1,\theta_2)&\to&\alpha A_0(x^1,x^2)\,,\\
 \theta_1 &\to& \frac{x^1}{\Sigma_1}\,,\\
 \theta_2 &\to& \frac{x^2}{\Sigma_2}\,,\\
 \tau &\to& \frac{x^0}{\Sigma}\,,
\end{eqnarray}
where $\alpha$ has mass dimension $-2$ and $\Sigma_1, \Sigma_2$ and
$\Sigma$ have mass dimension $-1$. Then, the action (\ref{D2}) is
rewritten by
\begin{eqnarray}
S_{1+1} &=&\frac{LT}{2}\frac{1}{\Sigma_1\Sigma_2\Sigma}
    \int dx^0\!\int_0^{2\pi\Sigma_1}\!\!dx^1\!
	\int_0^{2\pi\Sigma_2}\!\!dx^2\, \mbox{Tr}
	\Biggr[ (F_{\tau\theta_1})^2+(F_{\tau\theta_2})^2
	-(F_{\theta_1\theta_2})^2+(D_{\tau}X^m)^2\nn
 &&\hspace{25ex}-(D_{\theta_1}X^m)^2 -(D_{\theta_2}X^m)^2
    +\frac{\alpha^4}{2(2\pi L)^2}\,[\phi^{m},\phi^{n}]^2\Biggr],\\
 &&F_{\tau\theta_1}=\Sigma\alpha\p_0 A_1
    -\frac{L_1}{L}\,\Sigma_1\alpha\p_{1}A_0
	+i\frac{\alpha^2}{2\pi L}\,[A_0,A_1],\label{Fie1}\\
 &&F_{\tau\theta_2}=\Sigma\alpha\p_0 A_2
    -\frac{L_2}{L}\,\Sigma_2\alpha\p_{2}A_0
	+i\frac{\alpha^2}{2\pi L}\,[A_0,A_2],\label{Fie2}\\
 &&F_{\theta_1\theta_2}=\frac{1}{NL}L_1L_2 I
    +\frac{L_1}{L}\,\Sigma_1\alpha\p_{1} A_2
	-\frac{L_2}{L}\,\Sigma_2\alpha\p_{2}A_1
	+i\frac{\alpha^2}{2\pi L}\,[A_1,A_2],\label{Fie3}\\
 &&D_{\tau}X^m=\Sigma\alpha\p_0\phi^m
	+i\frac{\alpha^2}{2\pi L}\,[A_0,\phi^m],\\
 &&D_{\theta_1}X^m=\frac{L_1}{L}\,\Sigma_1\alpha\p_{1}\phi^m
	+i\frac{\alpha^2}{2\pi L}\,[A_1,\phi^m],\\
 &&D_{\theta_2}X^m=\frac{L_2}{L}\Sigma_2\alpha\p_{2} \phi^m
	+i\frac{\alpha^2}{2\pi L}\,[A_2,\phi^m].
\end{eqnarray}
In order to bring the field strength (\ref{Fie1})-(\ref{Fie3})
into the standard form (\ref{sta2}), we obtain the following relations,
\begin{eqnarray}
  \Sigma&=&\frac{\alpha}{2\pi L}\,,\\
  \Sigma_1&=&\frac{\alpha}{2\pi L_1}\,,\label{T-dual1}\\
  \Sigma_2&=&\frac{\alpha}{2\pi L_2}\,.\label{T-dual2}
\end{eqnarray}
{\it Eqs.(\ref{T-dual1}) and (\ref{T-dual2}) represent the T-duality
which relates the lengths $\Sigma_1$ and $\Sigma_2$ in the super
Yang-Mills theory and the lengths $L_1$ and $L_2$ in M-theory.}
{}From the D-brane viewpoint \cite{Sei,Sen}, this super Yang-Mills
theory is regarded as the low energy effective theory of $N$ D2-branes
and hence this is just a T-duality between D0-branes and D2-branes
\cite{Tay}.
We should stress, however, that we have obtained the same relations
from the membrane viewpoint.
Then, we have obtained the standard form of a bosonic part of
2+1-dimensional maximally supersymmetric $U(N)$ Yang-Mills theory with
constant magnetic flux,
\begin{eqnarray}
S_{1+1} &=&\frac{1}{2g_{YM}^2}\int dx^0\int_0^{2\pi\Sigma_1}\!\!dx^1\,
    \int_0^{2\pi\Sigma_2}\!\!dx^2\,\mbox{Tr}\Biggl[(F_{01})^2
	+(F_{02})^2 -(F_{12})^2+(D_{0}\phi^m)^2\nn
 &&\hspace{5cm}-(D_{1}\phi^m)^2 -(D_{2}\phi^m)^2
	+\frac{1}{2}[\phi^m,\phi^n]^2\Biggr], \\
 &&F_{01}=\p_0 A_1 -\p_{1}A_0 +i[A_0,A_1],\label{Fie1'}\\
  &&F_{02}=\p_0 A_2 -\p_{2}A_0+i[A_0,A_2],\label{Fie2'}\\
  &&F_{12}=\frac{1}{2\pi N\Sigma_1\Sigma_2}I+\p_{1} A_2 -\p_{2}A_1
	+i[A_1,A_2],\label{Fie3'}\\
 &&D_{0}\phi^m=\p_0 \phi^m +i[A_0,\phi^m],\\
 &&D_{1}\phi^m=\p_{1}\phi^m +i[A_1,\phi^m],\\
 &&D_{2}\phi^m=\p_{2} \phi^m +i[A_2,\phi^m],
\end{eqnarray}
with the boundary conditions,
\begin{eqnarray}
 A_0(x^1+2\pi\Sigma_1,x^2)&=&VA_0(x^1,x^2)V^{\dagger},\\
 A_1(x^1+2\pi\Sigma_1,x^2)&=&VA_1(x^1,x^2)V^{\dagger},\\
 A_2(x^1+2\pi\Sigma_1,x^2)&=&VA_2(x^1,x^2)V^{\dagger},\\
 \phi^m(x^1+2\pi\Sigma_1,x^2) &=&V\phi^m(x^1,x^2)V^{\dagger},\\
 A_0(x^1,x^2+2\pi\Sigma_2)&=&UA_0(x^1,x^2)U^{\dagger},\\
 A_1(x^1,x^2+2\pi\Sigma_2)&=&UA_1(x^1,x^2)U^{\dagger},\\
 A_2(x^1,x^2+2\pi\Sigma_2)&=&UA_2(x^1,x^2)U^{\dagger},\\
 \phi^m(x^1,x^2+2\pi\Sigma_2) &=&U\phi^m(x^1,x^2)U^{\dagger},
\end{eqnarray}
where $g_{YM}$ is the gauge coupling constant of mass dimension one
half, which is given by $g^2_{YM}=
(2\pi)^{-2}(\Sigma_1\Sigma_2)^{-1/2}(L_1L_2)^{-3/2}T^{-1}$.
We also define the dimensionless gauge coupling constant
$\tilde{g}_{YM}$ by
\begin{eqnarray}
 \tilde{g}_{YM}^2\equiv g_{YM}^2(2\pi\Sigma_12\pi\Sigma_2)^{1/2}
	=(2\pi)^{-1}(L_1L_2)^{-3/2}\,T^{-1}
	=2\pi\,\frac{l_{11}^3}{(L_1L_2)^{3/2}}\,.
\end{eqnarray}
This dimensionless gauge coupling constant exactly agrees with that
obtained in Ref.\cite{FHRS} including the numerical
constant.\footnote{Note that the parameters $\Sigma_1,\Sigma_2$ and
$L_1,L_2$ in Ref.\cite{FHRS} represent the circumferences but not the
radii.}
Note that in Refs.\cite{FHRS} the super Yang-Mills theory was regarded
as the low energy effective theory of D-branes in deriving such a
relation, while we have taken a different approach of matrix
regularization of supermembrane in this paper.
Furthermore, the constant magnetic flux $(2\pi
N\Sigma_1\Sigma_2)^{-1} I$ in eq.(\ref{Fie3'}) agrees with that
obtained in Refs.\cite{GRT,FHRS} including the numerical constant.

Finally, we comment on the relation to string. If we consider $X^9$ as
the 11th direction, this theory would be the light-cone type-IIB
superstring theory with the string coupling constant
$g^{IIB}_S=L_1/L_2$. The flip of $X^8$ and $X^9$ directions
corresponds to the S-duality in type-IIB superstring theory
\cite{Sch,Asp}.

%%%%%%%%%%%%%%%%%%%%%%%%%%%%%%%%%%
\section{Summary and discussion}\label{S:sum}
%%%%%%%%%%%%%%%%%%%%%%%%%%%%%%%%%%
In this paper, we have studied systematically matrix regularization
for the supermembrane in the light-cone gauge.
The regularization procedure is applicable for both unwrapped and
wrapped supermembranes and is summarized as the following mathematical
steps:
(i) {\it Introduce the noncommutativity on the space sheet of the
supermembrane, i.e., replace the product of functions on the space
sheet to the star product.}
(ii) {\it If possible, truncate the generators of the star-commutator
algebra in an algebraically consistent way.}
(iii) {\it Give a matrix representation of the (truncated)
star-commutator algebra.}
Following this procedure, we have deduced the $p$+1-dimensional
($p=0,1,2$) maximally supersymmetric $U(N)$ Yang-Mills theory from the
eleven-dimensional supermembrane in the light-cone gauge.
We have given the complete correspondence of the
super Yang-Mills theory to the eleven-dimensional supermembrane theory.
That is, in eqs.(\ref{mode1''})-(\ref{mode2''}),
(\ref{mode'1''})-(\ref{mode'3''}) and
(\ref{mode'''1})-(\ref{mode'''4}), the matrix elements in the super
Yang-Mills theory are determined by the Fourier coefficients in the
supermembrane theory.
We stress that we have never regarded the super Yang-Mills theory as
the low energy effective theory of D-branes in this paper.
Nevertheless, we have obtained the T-duality relations which relates
the lengths in the super Yang-Mills theory and M-theory and the same
dimensionless gauge coupling constant as that obtained in
Ref.\cite{FHRS} in which the super Yang-Mills theory is regarded as
the low energy effective theory of D-branes.
Thus our results gives a consistency check of two kinds of
profiles of the super Yang-Mills theory, i.e., one is the matrix
regularized theory of the eleven-dimensional light-cone supermembrane
and the other is the low energy effective theory of D-branes.

Finally, we discuss the derivation of 3+1-dimensional maximally
supersymmetric $U(N)$ Yang-Mills theory from our point of view (the
membrane viewpoint).
Naively, it seems that it is obtained by the matrix regularization of
the wrapped supermembrane on $R^8\times T^3$.
However, we shall come up against a problem immediately.
If we take $X^9,X^8$ and $X^7$ as coordinates of three cycles of
$T^3$, the supermembrane can wrap on $T^3$ in three ways, i.e.,
wrapping around $X^9$-$X^8$ surface, $X^8$-$X^7$ surface and
$X^7$-$X^9$ surface of $T^3$.
In the language of the 3+1-dimensional super Yang-Mills theory,
three ways of the wrapping correspond to three components
of the magnetic flux $F_{12},F_{23},F_{31}$ \cite{Sus2,GRT,FHRS}.
Thus the 3+1-dimensional super Yang-Mills theory
includes three ways of the wrapping simultaneously.
On the other hand, eq.(\ref{MLCgauge}) is
the action of the first-quantized supermembrane
theory in the light-cone gauge, i.e., the action of the one-body
system.\footnote{It is widely appreciated that the first-quantized
supermembrane theory through its continuous spectrum \cite{dWLN} is
capable of describing multiple supermembranes. This picture may solve
the difficulty in the derivation of the 3+1-dimensional super Yang-Mills
theory through the matrix regularization of the wrapped supermembrane
on $R^8\times T^3$.}
Hence the matrix-regularized action cannot represent three ways of the
wrapping simultaneously in the present formulation. This deserves to
be studied further.

%%%%%%%%%%%%%%%%%%%%%%%%%%%%%%%%%%%
\vspace{\baselineskip}

\noindent{\bf Acknowledgments:}
This work is supported in part by MEXT Grant-in-Aid for
the Scientific Research \#13135212 (S.U.).

%%%%%%%%%%%%%%%%%%%%%%%%%%%%%%%%%

%%%%%%%%%%%%%%%%%%%%%%%%%%%%%%%%%
\end{document}